\documentclass[conference]{IEEEtran} %

\IEEEoverridecommandlockouts
\usepackage{cite}
\usepackage{amsmath,amssymb,amsfonts}
\usepackage{algorithmic}
\usepackage{graphicx}
\usepackage{textcomp}
\usepackage{xcolor}
\usepackage{multirow}
\usepackage{comment}
\def\BibTeX{{\rm B\kern-.05em{\sc i\kern-.025em b}\kern-.08em
    T\kern-.1667em\lower.7ex\hbox{E}\kern-.125emX}}

\begin{document}

\title{Characterizing 5G User Throughput via Uncertainty Modeling and Crowdsourced Measurements\\

\thanks{This work has been partially supported by the project 5GPERFORMANCE (TSI-064200-2023-0005), funded by the UNICO I+D 6G program of the Spanish Ministry of Digital Transformation. Zoraida Frias acknowledges funding from UPM’s \textit{Doctores Emergentes} program (24-ED3B0Y-100-JOQZ29), supported by the Region of Madrid. Javier Albert-Smet acknowledges financial support from the 5G+RADIOLAB project (TSI-064100-2022-8), funded by the Spanish Ministry of Economy and Finance.}
}

\author{
\IEEEauthorblockN{Javier Albert-Smet\IEEEauthorrefmark{1}, 
Zoraida Frias\IEEEauthorrefmark{1},
Luis Mendo\IEEEauthorrefmark{1},
Sergio Melones\IEEEauthorrefmark{1}\IEEEauthorrefmark{2},
Eduardo Yraola\IEEEauthorrefmark{2}}
\IEEEauthorblockA{\IEEEauthorrefmark{1}\textit{Universidad Politécnica de Madrid}. Madrid, Spain}
\IEEEauthorblockA{\{javier.albert.smet, zoraida.frias, luis.mendo\}@upm.es}
\IEEEauthorblockA{\IEEEauthorrefmark{2}\textit{Weplan Analytics}. Madrid, Spain}
\IEEEauthorblockA{\{sergio.melones, eduardo.yraola\}@weplananalytics.com}
}

\maketitle

\begin{abstract}

Characterizing application-layer user throughput in next-generation networks is increasingly challenging as the higher capacity of the 5G Radio Access Network (RAN) shifts connectivity bottlenecks towards deeper parts of the network. Traditional methods, such as drive tests and operator equipment counters, are costly, limited, or fail to capture end-to-end (E2E) Quality of Service (QoS) and its variability. In this work, we leverage large-scale crowdsourced measurements—including E2E, radio, contextual and network deployment features collected by the user equipment (UE)—to propose an uncertainty-aware and explainable approach for downlink user throughput estimation. We first validate prior 4G methods, improving $R^2$ by 8.7\%, and then extend them to 5G~NSA and 5G~SA, providing the first benchmarks for 5G crowdsourced datasets. To address the variability of throughput, we apply NGBoost, a model that outputs both point estimates and calibrated confidence intervals, representing its first use in the field of computer communications. Finally, we use the proposed model to analyze the evolution from 4G to 5G~SA, and show that throughput bottlenecks move from the RAN to transport and service layers, as seen by E2E metrics gaining importance over radio-related features.


\end{abstract}


\section{Introduction}
\label{sec:intro}

Over the past two decades, the widespread adoption of mobile broadband networks has intensified both user and industry demands for reliable and predictable Quality of Service (QoS) to support effective network management and optimization. Among QoS indicators, user throughput is a primary concern for bandwidth-intensive applications such as media streaming and large file transfers. However, obtaining a realistic characterization of achievable user data rates in next-generation networks is increasingly challenging due to fluctuating radio conditions, complex deployments with continuous optimization, and highly volatile user demands.

Traditional approaches to monitoring throughput, including operator-side counters and drive-test campaigns, provide only partial observability of user-experienced QoS. Drive-test campaigns are costly, geographically restricted, and quickly outdated, while operator counters are limited to network-layer measurements. Consequently, end-to-end (E2E) visibility is constrained, and network-wide assessments of user QoS cannot be achieved at scale.

In this context, crowdsourced measurements—large-scale data collected anonymously from mobile users—have emerged as a promising alternative that avoids the high costs and limited coverage of traditional drive-test campaigns. Passive radio metrics (e.g. RSRP, RSRQ) and active tests (e.g. throughput and latency tests) obtained at the user equipment (UE) side can be aggregated to generate fine-grained spatio-temporal QoS maps, whose strength lies in both the volume of data collected and the diversity of performance indicators provided. This large volume and richness of data enables the use of machine learning (ML) to characterize QoS metrics, such as expected user throughput. 

This paper formulates the problem of estimating 5G application-layer downlink user throughput from E2E, radio, contextual, and network deployment features collected at the UE through crowdsourced measurements. We propose an uncertainty-aware and explainable approach to throughput estimation that aims to pinpoint the network side factors that affect user download speeds and their variability from an E2E perspective. Our contributions are as follows:

\begin{itemize}
    \item We validate throughput estimation methods previously proposed for 4G crowdsourced data \cite{Ghasemi2019}, achieving improved accuracy in our dataset, and extend the methodology to 5G Non-Stand Alone (NSA) and 5G Stand Alone (SA). This provides the first benchmark for throughput estimation in 5G crowdsourced datasets, while also demonstrating the limitations of current throughput characterization approaches in terms of input observability and QoS variability (see Subsection~\ref{sec:benchmarking}).
    \item We propose an uncertainty-aware approach for characterizing downlink user throughput in high-variability scenarios, such as 5G~SA deployments. The approach, based on NGBoost, outputs both central estimates and calibrated confidence intervals, representing the first application of this algorithm to the field of computer communications. It achieves an interpretability of central estimates comparable to XGBoost, while additionally enabling uncertainty explainability (see Subsection~\ref{sec:uncertainty_evaluation}).
    \item We leverage the proposed approach to analyze the evolution from legacy 4G to 5G~SA networks, revealing that throughput bottlenecks shift from the Radio Access Network (RAN) towards transport and service layers. In this transition, E2E features become more influential than radio-related metrics for the characterization of user throughput, and RSRP overtakes RSRQ as the dominant radio indicator. Furthermore, time-of-day effects are shown to diminish in 5G~SA, underscoring the capacity of next-generation deployments to manage fluctuations in user demand (see Subsection~\ref{sec:results_xai}).
\end{itemize}

\section{Background and Related Work}
\label{sec:bg}

Several studies have investigated throughput modeling\footnote{In the context of this work, modeling serves as an umbrella term encompassing both estimation (deriving expected throughput from observed data at a given point in time) and prediction (forecasting future throughput values). Characterization refers to throughput modeling with an added focus on explainability.}, ranging from large-scale estimation of expected data rates for QoS optimization to time-series prediction of channel bandwidth for resource management. The most relevant prior work is \cite{Ghasemi2019}, which addresses user throughput characterization from 4G crowdsourced data. However, the authors report a relatively low explained variance ($R^2 = 0.537$). Their feature contribution analysis indicates that this limitation arises from the inherent variability of large-scale user measurements, driven by fluctuating network conditions, UE heterogeneity, and diverse deployment scenarios.  

Since then, throughput characterization from crowdsourced data has seen limited progress and has not been extended to 5G networks, likely due to the modest results obtained. This work revisits the problem to show that it remains both relevant and approachable when addressed with richer data and uncertainty-aware methods. To this end, related studies can be grouped into two complementary directions, both unified in our approach: enhancing the observability of key throughput drivers (Section~\ref{sec:bg_observability}) and quantifying the inherent variability in the data (Section~\ref{sec:bg_uncertainty}).

\subsection{Improving Observability in Throughput Estimation}
\label{sec:bg_observability}

Input observability for throughput estimation has been addressed by several studies. In \cite{Ghasemi2019}, using a 4G crowdsourced dataset from OpenSignal, estimation accuracy improved with the incremental inclusion of radio, context, and deployment features, showing that performance benefits substantially from data beyond radio metrics alone. Echoing these findings, the authors of \cite{Zhohov2021} combined vendor-side information with UE radio measurements for 4G uplink throughput prediction. Their feature importance analysis confirmed that radio metrics can be combined with network side features to improve estimation accuracy. 

Similarly, \cite{Paul2022} leveraged M-Lab and Ookla datasets to account for factors such as user subscription plans, device types, test vendors, and time-of-day effects, further demonstrating that richer feature sets not only enhance accuracy but also improve the interpretability of active speed test results. 

Finally, shifting the focus from feature inclusion to modeling approaches, \cite{Minovski2023} compared Random Forests, Support Vector Regressors, Neural Networks, and Gradient Boosting for throughput prediction, showing that these algorithms achieve broadly comparable accuracies. 

Consequently, these studies suggest that throughput modeling is primarily data-limited: Improved observability consistently enhances performance, while alternative modeling strategies deliver comparable results.

\subsection{Towards Uncertainty-Aware Measurements}
\label{sec:bg_uncertainty}

While greater observability of input features improves modeling accuracy, in practice full network visibility is limited, since data access is fragmented across stakeholders. QoS estimates are therefore inherently non-deterministic (e.g. it is impossible to access complete information to determine user throughput with absolute accuracy). This motivates uncertainty-aware approaches that explicitly capture this variability.

Crowdsourced data enable the capture of network QoS variations, as analyzed in \cite{Wamser2021}, where measurements are treated as a sampling process. The authors assess whether sufficient measurements exist for reliable statistics (e.g., mean throughput), and locate regions in France where data density ensures accuracy within 100 kbps. However, the approach remains restricted to spatiotemporal statistics and does not incorporate modeling or additional contextual, deployment, radio, or end-to-end features that could enable a more comprehensive characterization.

Beyond mobile network QoS monitoring, uncertainty quantification has a longer history in communication systems. Early work by \cite{Jain2005} introduced uncertainty measurements for predicting the available bandwidth of a connection, providing an initial approach to quantifying and modeling QoS variability in wired networks. More recently, uncertainty quantification has been explored in wireless systems, including positioning \cite{Tedeschini2024}, model placement \cite{Huang2024}, reliable decision-making \cite{Zecchin2023}, and uncertainty-aware framework design \cite{Wang2024}. Together, these studies highlight the growing importance of incorporating uncertainty-awareness into modeling.

\section{Methodology and Experimental Setup}
\label{sec:methods}
This Section describes the input dataset, the applied data transformations, the modeling approach, and the evaluation of the outputs. The modeling code is publicly available \cite{AlbertSmet2025}.

\subsection{Dataset Description}
\label{sec:dataset_description}

Our dataset comprises over a quarter million downlink throughput speed test samples collected from the three major mobile network operators in Spain.
The data was collected by Weplan Analytics, a global crowdsourced data provider specializing in mobile network analytics and benchmarking. Measurements were gathered over a four-month period, from May to September 2025.

Each sample in the dataset contains radio metrics (RSRP, RSRQ, SINR, and timing advance) \cite{Johnson2019}, E2E measurements (latency, jitter, Time-to-First-Byte and packet loss) \cite{ITU-T-Y1540}, deployment (frequency band and carrier) and contextual information (time and day of the week), as well as the measured downlink throughput from a file download in 4G, 5G~NSA and 5G~SA networks.

Weplan Analytics has a proprietary methodology based on the collection of throughput speed test, radio metrics and latency data through Android applications, deployed across millions of devices globally. This methodology facilitates measurements of available bandwidth and latency. All data is collected anonymously, with no personally identifiable information acquired, and location permissions used only when granted, ensuring compliance with privacy regulations \cite{WePlanAnalytics}. 

\subsection{Feature Engineering}

This subsection describes the processing of input data required to prepare features for model ingestion.

\subsubsection{Training, validation, and testing dataset split} 

To avoid data leakage across training, validation, and testing, a temporal split was applied. The most recent three weeks of data were withheld: one week for validation and two weeks for testing. The remaining \(\sim 70\%\) of samples were used to train models for each radio access technology. This chronological partitioning more faithfully represents deployment scenarios where models must generalize to future, unseen conditions.

\subsubsection{Input Features} 
The raw features described in Subsection~\ref{sec:dataset_description} were pre-processed to enable ingestion by tree-based learning algorithms. Because decision-tree ensembles (e.g., gradient boosting) partition the feature space via threshold splits, no explicit normalization or clipping was required. 

Missing entries were left unaltered in the dataset. Tree-based methods can handle them natively, and their occurrence can provide informative signals that contribute to uncertainty estimation.

Finally, contextual features were derived from the measurement timestamp, including time-of-day and day-of-week indicators. Each of these, as well as the network operator identity (carrier), was numerically encoded separately for model ingestion. All input features were retained irrespective of their mutual correlations, as the primary objective is to characterize the network and identify the underlying conditions influencing throughput performance.

\subsubsection{Target Variable} 

Downlink throughput, measured in \textit{kbps}, was log-transformed using $\log(1+y)$ scaling to account for long-tailed distributions and to avoid undefined values at zero. The transformed values were then normalized using the training dataset’s mean and variance. This preprocessing facilitates comparability across studies and aligns the reported performance metrics with the baseline established in \cite{Ghasemi2019}. 

\subsection{Modeling}

As seen in Section~\ref{sec:bg}, throughput estimation is data-limited. We therefore restricted hyperparameter tuning to training length parameters (e.g., number of estimators), since model hyperparameters had little effect on performance. Our main focus was instead on the training paradigm—whether the model produces point estimates or full predictive distributions—implementing two gradient boosting approaches: a baseline and an uncertainty-aware model.

\subsubsection{Baseline Model}
XGBoost is used as a baseline to replicate point-estimate results reported in the literature for our data \cite{Ghasemi2019}. This regressor was configured with the default parameters—maximum tree depth of 6, and a learning rate of 0.05. The number of estimators is capped at 1,000 to allow sufficient calibration, with early stopping triggered after 100 consecutive rounds without improvement in Root Mean Squared Error (RMSE).

\subsubsection{Uncertainty-Aware Model}
The second model used is NGBoost, an open-source framework that extends gradient boosting to predictive uncertainty estimation by outputting full probabilistic predictions for real-valued targets \cite{Duan2020}. To the best of our knowledge, this represents the first application of NGBoost in the field of computer communications, extending its adoption beyond the diverse domains where it has previously been applied.

The NGBoost regressor was configured to model the target variable as a Normal distribution, optimized using the negative log-likelihood (NLL) score. As in the baseline model, a maximum 1,000 estimators was established, a learning rate of 0.05 and early stopping after 100 non-improving rounds. The base learner was the default decision tree regressor with a maximum depth of 3. Compared to XGBoost, NGBoost typically employs shallower trees, as lower maximum depth reduces variance and stabilizes the estimation of predictive distributions.

\subsection{Model Evaluation}
\subsubsection{Point Estimation Accuracy Evaluation}

Performance was assessed using the mean absolute error (MAE) and the RMSE, reported both in standardized logarithmic units and in natural unscaled units. The RMSE complements the MAE as it weights large errors more heavily. In addition, the coefficient of determination ($R^2$) is also reported to quantify the explained variance. These three complementary metrics were selected as they are standard in ML model evaluation.

\subsubsection{Uncertainty Evaluation}
The quality of NGBoost's output distributions was evaluated in terms of the continuous ranked probability score (CRPS) and the calibration curves. CRPS extends MAE to probabilistic estimates by measuring the discrepancy between the predicted cumulative distribution and the observed outcome. Calibration, in turn, assesses the agreement between predicted confidence intervals and their empirical coverage (i.e., the proportion of throughput measurements falling within the predicted intervals). To quantify this, we used the Calibration Area Under the Curve (C-AUC), defined as the area between the empirical calibration curve and the diagonal of perfect calibration, computed via the trapezoidal rule over the absolute difference between both curves.

\subsection{Explainability Analysis}
The widespread SHapley Additive exPlanations (SHAP) framework is used to evaluate feature importance for the different models and the different radio access technologies. Both models are evaluated through the TreeExplainer, a fast and exact method to estimate SHAP values for tree models and ensembles of trees.

To enable consistent comparison of feature rankings across radio access technologies and between the baseline and uncertainty-aware models, SHAP values were normalized by dividing each by the maximum value obtained within the corresponding model.

SHAP values for the NGBoost model were computed for both the mean and standard deviation of the predictive distribution. This facilitates interpretation of the network conditions that influence not only throughput levels but also their variability.

\section{Results and Discussion}

The results are organized to address the main contributions outlined in the Section~\ref{sec:intro}. The following three subsections present, respectively, the benchmarking of throughput estimation methods, the evaluation of uncertainty-aware modeling, and the explainability analysis across network generations.

\subsection{Point Estimation Accuracy Benchmarking}
\label{sec:benchmarking}

\begin{table}[htbp] \caption{Performance Metrics for the Models and Network Types} \begin{center} \scriptsize \begin{tabular}{|c|c|c|c|c|} \hline \textbf{Model} & \textbf{MAE} & \textbf{RMSE} & \textbf{$R^2$} & \textbf{CRPS} \\ \cline{2-3} & \textit{log norm / kbps} & \textit{log norm / kbps} & & \\ \hline 4G XGBoost & 0.4677 / 13875 & 0.6362 / 21283 & 0.5837 & 0.4677 \\ \hline 4G NGBoost & 0.4704 / 14004 & 0.6374 / 21451 & 0.5821 & 0.3371 \\ \hline 5G~NSA XGBoost & 0.5297 / 42598 & 0.7068 / 57530 & 0.4624 & 0.5297 \\ \hline 5G~NSA NGBoost & 0.5345 / 43228 & 0.7101 / 58398 & 0.4574 & 0.3777 \\ \hline 5G~SA XGBoost & 0.5488 / 49458 & 0.8305 / 63500 & 0.3987 & 0.5488 \\ \hline 5G~SA NGBoost & 0.5461 / 49723 & 0.8238 / 63870 & 0.4085 & 0.3980 \\ \hline \end{tabular} \label{tab:pointwise_accuracy_metrics} \end{center} \end{table}

\begin{figure}
    \centering
    \includegraphics[width=\linewidth]{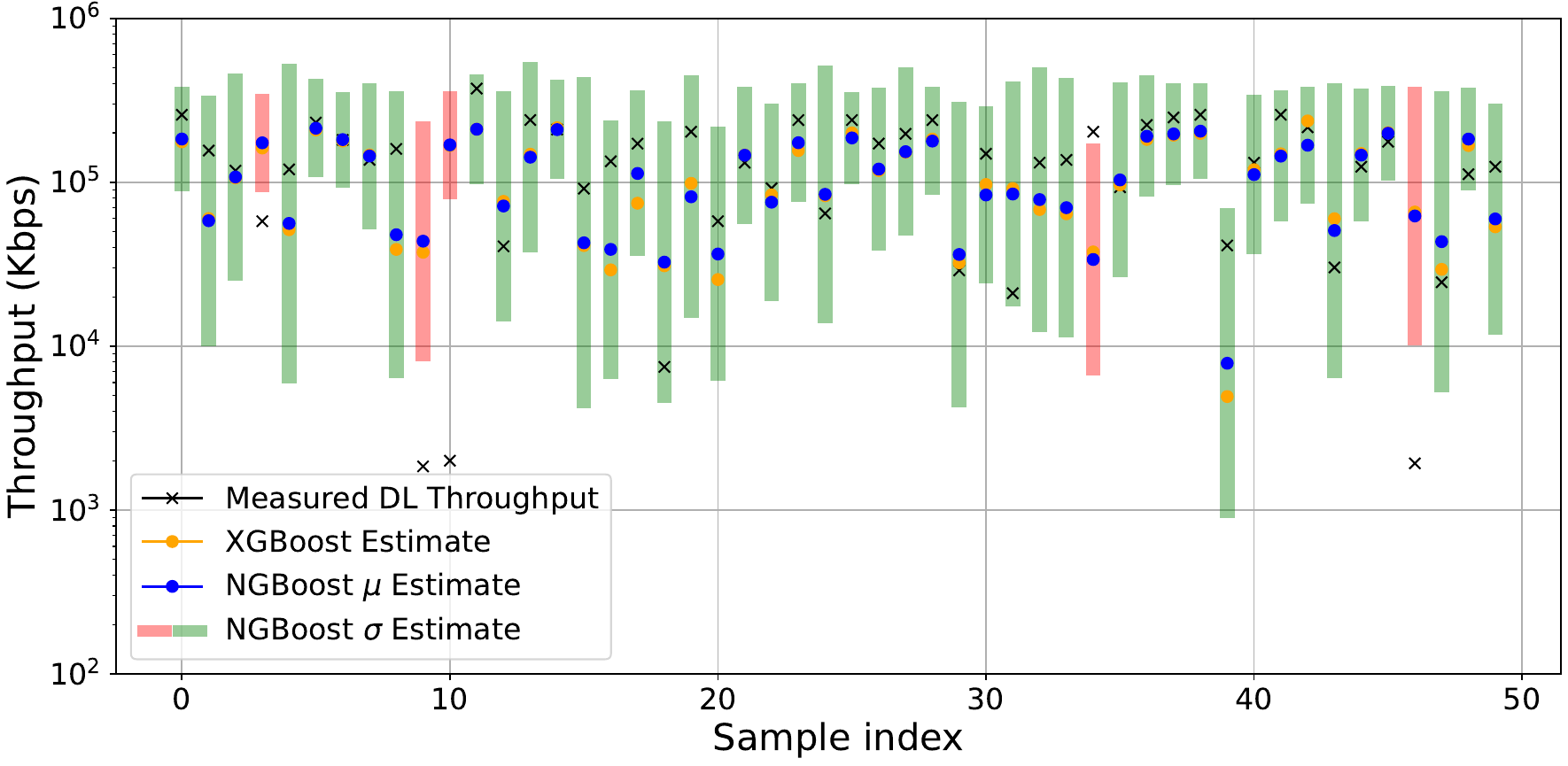}
    \caption{NGBoost 95\% confidence intervals for 50 random 5G~SA samples. Dots show XGBoost point estimates (orange) and NGBoost means (blue); intervals are green when containing the true value (black cross) and red otherwise.}
    \label{fig:5g-ci}
\end{figure}

Table~\ref{tab:pointwise_accuracy_metrics} presents the evaluation metrics from Section~\ref{sec:methods}, and Fig.~\ref{fig:5g-ci} provides a visual reference of 50 random outputs of XGBoost and NGBoost for 5G~SA, including confidence intervals and the point estimates.

When comparing the baseline XGBoost model on the 4G network to the prior benchmark \cite{Ghasemi2019}, all metrics improve, with reductions of 20\% in MAE and RMSE and an increase of 8.7\% in $R^2$. This enhancement likely stems from incorporating E2E metrics in the training (see Section \ref{sec:results_xai}), which reduces point estimation error metrics and increases the explainable variance ratio, $R^2$. 

Moreover, the point-estimate accuracy of XGBoost can be directly compared with the mean predictions of NGBoost. For 4G and 5G~NSA, XGBoost achieves slightly better results in terms of MAE and RMSE, whereas in 5G~SA it is marginally outperformed by NGBoost. These differences remain small, showing that the uncertainty-aware model can deliver state-of-the-art point estimates across technologies, while also establishing a benchmark for throughput estimation from crowdsourced data in 5G~SA networks. 

Although both models achieve reasonable accuracy, the coefficient of determination ($R^2$) remains relatively low, particularly in 5G networks. This indicates that point estimators struggle to capture the explainable variance inherent to next generation deployments with the available feature set. This reinforces the conclusion in the literature that throughput estimation is primarily data-limited. 

While greater observability of input features can improve modeling accuracy—as shown by the inclusion of E2E features compared to \cite{Ghasemi2019}—the potential for reducing RMSE or MAE remains constrained by the information available in the features of our dataset. The modeling paradigm must therefore shift from producing highly accurate point estimates to evaluating how effectively the input features capture the bounds of output variability, as illustrated in Fig.~\ref{fig:5g-ci}.

\subsection{Uncertainty-Aware Model Evaluation}
\label{sec:uncertainty_evaluation}

\begin{figure}
    \centering
    \includegraphics[width=0.75\linewidth]{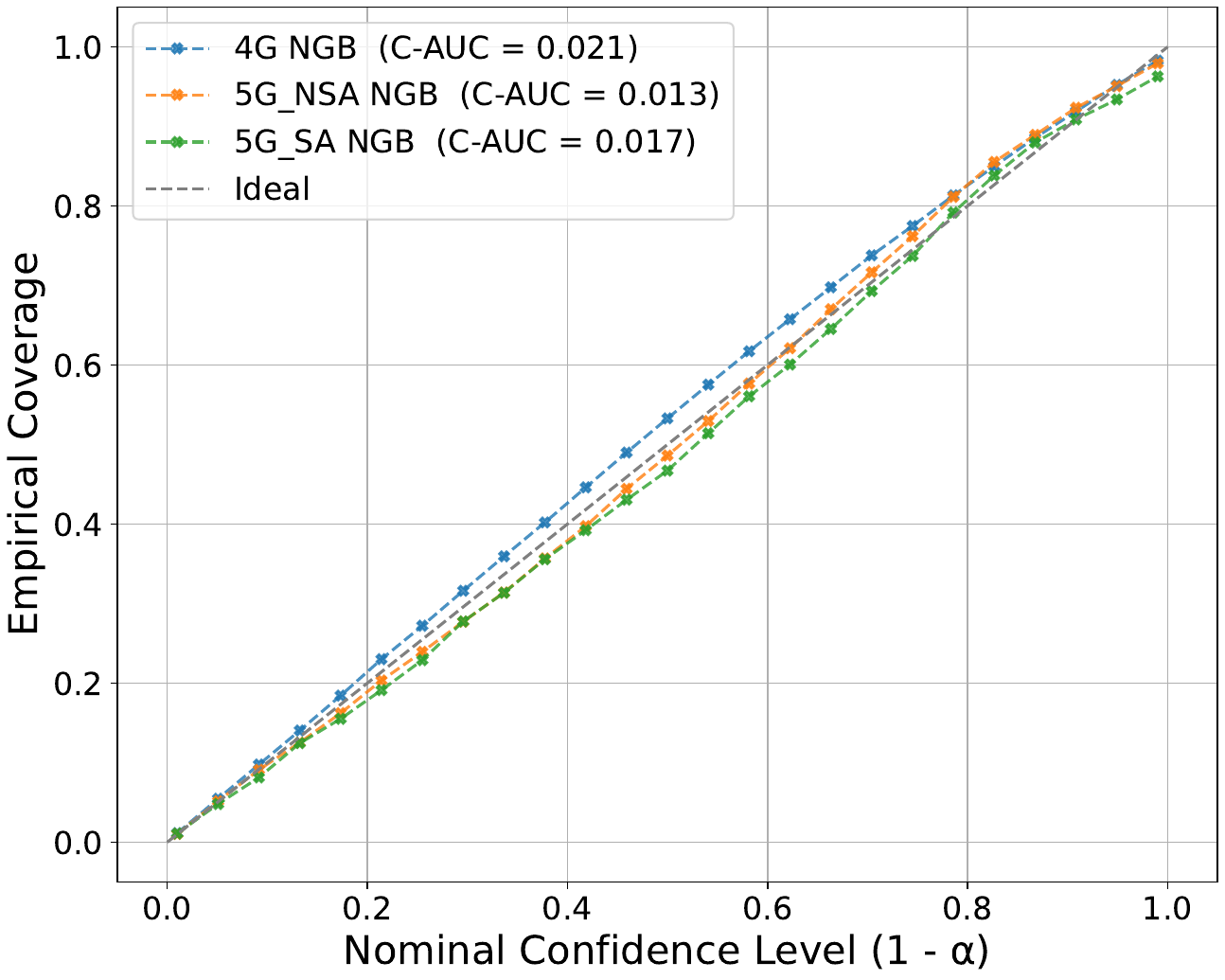}
    \caption{Reliability diagram with the calibration curves for the NGBoost model for the different radio access technologies' testing sets, where $\alpha$ is the miscoverage level. The C-AUC is indicated in the legend. }
    \label{fig:cal-curves}
\end{figure}

The ability of NGBoost to capture uncertainty is first assessed through the CRPS score reported in Table~\ref{tab:pointwise_accuracy_metrics}. A CRPS lower than the MAE for the uncertainty-aware model indicates that the distributional scoring rule penalizes less than a strict pointwise metric, reflecting that the models' uncertainty estimates provide meaningful information beyond the mean prediction. A more detailed view of this calibration behavior on the test set is shown in Fig.~\ref{fig:cal-curves}.

NGBoost’s confidence intervals closely follow the ideal calibration curve. For 5G technologies, the C-AUC is within 2\% of the maximum value (0.5), indicating good calibration—for instance, 93.4\% of the data falls within the predicted 95\% confidence interval in 5G~SA. In 4G, the C-AUC is slightly higher, reflecting a mild but consistent underconfidence. 

Overall, the model produces well-calibrated confidence intervals and adapts their width to input-dependent variability, establishing it as a suitable choice for uncertainty-aware throughput modeling.

\subsection{Insights on the Evolution of Throughput Characterization}
\label{sec:results_xai}

\begin{figure}
    \centering
    \includegraphics[width=\linewidth]{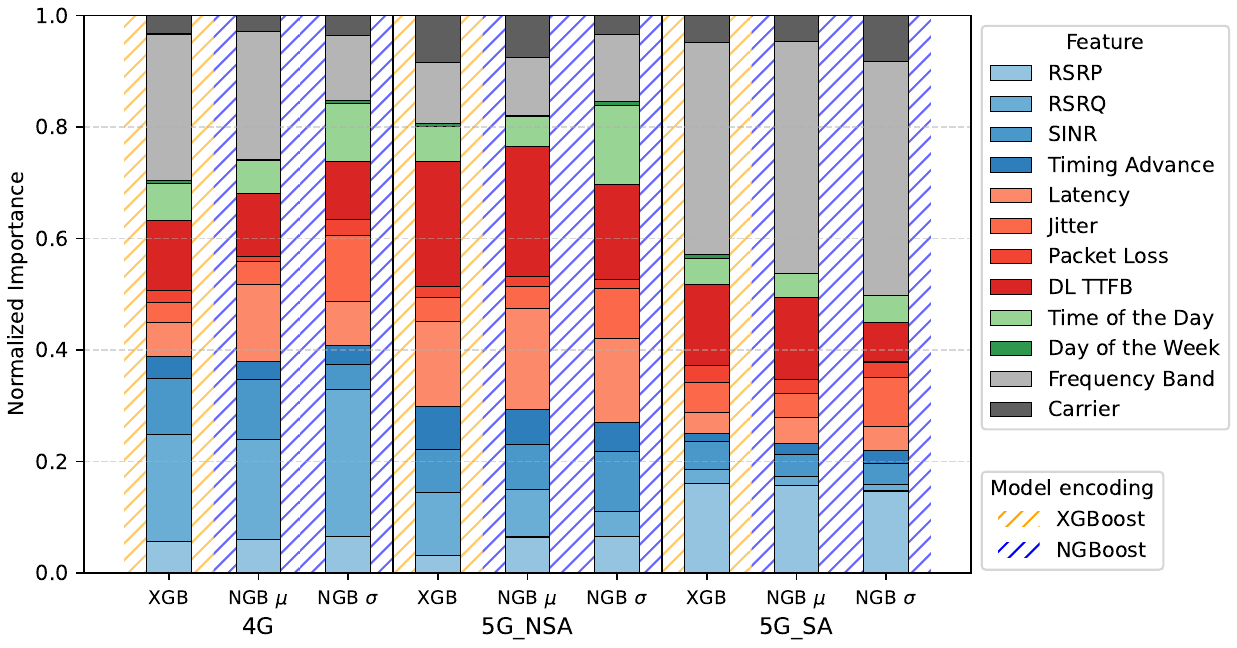}
    \caption{Average feature importance plot for the XGBoost and NGBoost models for 4G, 5G~NSA and 5G~SA. Feature categories are color-coded: radio (blue), E2E (red), contextual (green), and deployment (gray).}
    \label{fig:xai_models}
\end{figure}

Having established a reliable uncertainty-aware model for throughput estimation, we now leverage it to interpret the E2E network factors that drive application-layer QoS by identifying the dominant contributors to throughput performance, as illustrated in Fig.~\ref{fig:xai_models}.

From a modeling perspective, the normalized feature weights of point estimates XGBoost and NGBoost, $\bar{x}$, are practically equivalent, indicating that the uncertainty-aware model preserves a point estimate interpretability comparable to the baseline. The minor differences appear mainly in latency and jitter, which can be explained by NGBoost’s variance term, $\sigma$, learning that jitter is a primary source of uncertainty, while the mean term captures the central tendency associated with latency. This highlights the model’s ability to disentangle factors influencing central tendency and variability—an aspect not achievable with point estimates alone.

When applying the uncertainty-aware model to assess the main feature groups influencing user downlink throughput, a notable reduction in the importance of radio-related features is observed in the evolution from 4G to 5G. When computing the ratio of the summed feature importances of E2E variables to those of radio variables, the value increases from 0.79 in 4G to 1.12 in 5G~SA. This reflects that E2E metrics outweigh radio-related metrics in 5G for throughput estimation, suggesting that performance bottlenecks are moving away from the RAN and toward the backhaul, core, and service infrastructure in next-generation deployments.

Furthermore, when focusing solely on radio-related features, RSRQ—the dominant factor in 4G, accounting for 47\% of radio feature explainability—is surpassed by RSRP in 5G~SA, which accounts for 68\%. This indicates that New Radio (NR) is currently not capacity-limited, as in Long Term Evolution (LTE), but rather coverage-limited. This shift is consistent with the introduction of higher-frequency spectrum in next-generation deployments, which provides greater bandwidth but suffers from higher path loss.

Moreover, the main contextual feature, time of day, acquires different importance in the mean and variance explanations. It primarily accounts for variability in throughput, underscoring the model’s ability to attribute part of the uncertainty to traffic patterns indirectly encoded in this variable. Notably, its contribution to throughput variability decreases in 5G~SA networks, suggesting that next-generation deployments are more suited to cope with fluctuations in user data demand through improvements such as load balancing and increased network capacity. The fact that, at present, 5G~SA networks appear coverage-limited and less affected by traffic variability suggests that current deployments may be over-provisioned relative to throughput demands at this stage.

When analyzing the deployment features, it is worth noting that the frequency band maintains a clear dominant influence in 5G~SA. This result is expected since the 3500 MHz was introduced in 5G deployments in Spain, and conditions the maximum bandwidth available for the connection. However, in 5G~NSA networks, the frequency band reported by the UEs is the one of the LTE anchor (primary cell), which explains the lower influence of this deployment feature. In this case, the model compensates for the lack of secondary cell information by assigning greater weight to the E2E features. The model reveals that QoS management could be improved by incorporating neighboring cell information, specifically increasing observability targeting dual connectivity and carrier aggregation features.

\begin{figure}
    \centering
    \includegraphics[width=\linewidth]{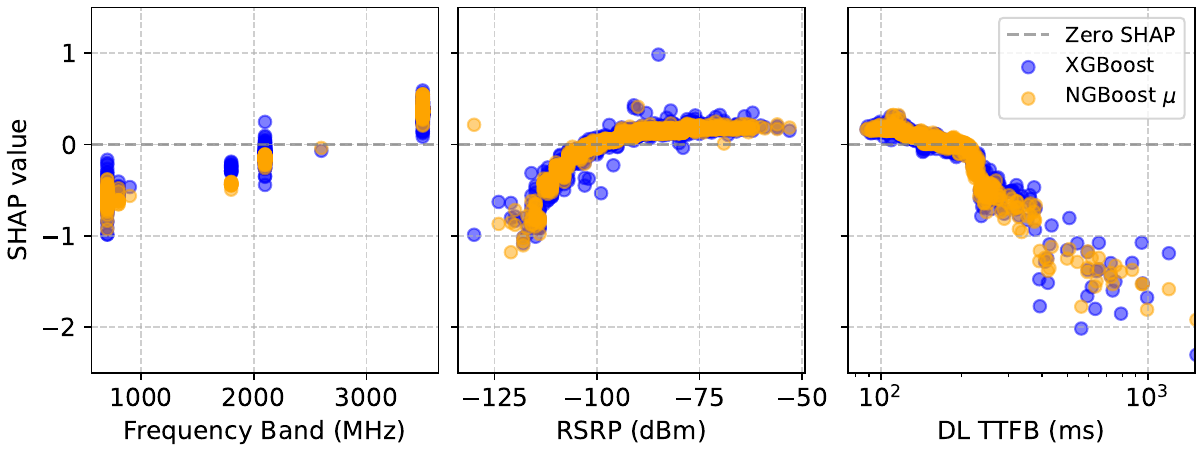}
    \caption{SHAP vs feature value plots of XGBoost and NGBoost for the three most important inputs in 5G~SA networks.}
    \label{fig:shap_spread}
\end{figure}

Fig.~\ref{fig:shap_spread} compares SHAP values from the baseline and uncertainty-aware models, complementing Fig.~\ref{fig:xai_models}. The main difference is that NGBoost, through the use of natural gradients, exhibits more stable training dynamics and a better fit, reflected in a lower spread of SHAP values. The figure also shows that the importance of RSRP decreases significantly above –100~dBm, as well as Time-to-First-Byte (TTFB) values below 200~ms having limited impact. This suggests that as long as application-layer latency remains above this threshold and the received radio power is sufficient, the achievable throughput is largely unaffected.

\section{Conclusions}

This work proposes an uncertainty-aware and explainable approach to user downlink throughput characterization based on large-scale crowdsourced measurements of 5G networks. By incorporating E2E, radio, contextual and deployment features collected by the UE, the approach provides point predictions with calibrated confidence intervals, together with both central and uncertainty interpretability for QoS-aware network management. 

The results highlight the increasing relevance of E2E measurements in next-generation deployments and underscore the need for greater network observability and explicit uncertainty characterization in data-limited QoS estimation tasks. While increasing the observability of input features can improve modeling accuracy, in practice it is bounded: data is fragmented across stakeholders, and economic incentives often prevent its full disclosure. As a result, there will always remain irreducible uncertainty to be captured in throughput modeling, stemming from unobserved factors that cannot be captured in the available data.

Compared with prior studies on 4G, our framework improves RMSE and MAE by 20\% and $R^2$ by 8.7\%, while delivering the first reference models for throughput estimation in 5G~NSA and 5G~SA. To the best of our knowledge, this represents the first uncertainty-aware QoS characterization at scale in mobile networks.

From a modeling perspective, we showed that XGBoost and NGBoost achieve comparable performance in point estimation and interpretability. Beyond this, NGBoost provides reliable confidence intervals and more stable learning dynamics, modeling the data with reduced variance. These properties make it well suited for throughput estimation with uncertainty, as it can capture finer distinctions between latency and jitter.

The explainability analysis revealed structural differences across different network technologies. E2E metrics surpass radio-related features, indicating that the performance bottleneck has shifted away from the RAN. Among radio variables, RSRP emerges as the dominant factor for estimating throuhgput, suggesting that 5G networks are, as for now, coverage-limited rather than capcity-limited. This is further supported by the reduced relevance of time-of-day effects in 5G throughput variability, reflecting the greater resource availability and improved adaptability to traffic fluctuations of next-generation deployments.

Finally, in future work, the proposed approach can inform where increased observability would most benefit QoS management. For instance, future studies can incorporate information from secondary cells (e.g. dual-connectivity features or total connection bandwidth) to enhance throughput characterization in 5G~NSA deployments, currently limited to primary cell measurements in this work. Similarly, improved collection of application-layer variability metrics is recommended, as bottlenecks increasingly emerge at this layer.  Moreover, research can extend to characterizing uplink throughput or other QoS metrics. The modeling framework could also benefit from further hyperparameter optimization to mitigate model underconfidence, or advanced feature engineering for improving accuracy. Alternative uncertainty quantification techniques, such as Bayesian Neural Networks or Gaussian Processes, could be explored for comparison. Finally, the explainability analysis, which relied on mean SHAP values, could be extended to the full distribution of SHAP contributions (e.g. median, negative, and tail values) to provide deeper insights into how input features shape both throughput and its variability.

\bibliographystyle{IEEEtran}   
\bibliography{references}      

\end{document}